\documentclass[%
 reprint,
 amsmath,amssymb,
 aps,
 prb,
]{revtex4-1}

\usepackage[dvips]{graphicx}
\usepackage[dvips]{color}
\usepackage[dvips]{hyperref}
\hypersetup{colorlinks=true, linkcolor=blue, urlcolor=blue, citecolor=blue}
\usepackage{breakurl}
\usepackage{dcolumn}
\usepackage{amsmath} \usepackage{amssymb}
\usepackage{accents} \usepackage{mathrsfs} \usepackage{setspace}
\usepackage{color}
\usepackage{tcolorbox}
\definecolor{dgreen}{RGB}{00, 120, 00} \definecolor{dblue}{RGB}{00, 00, 180}
\definecolor{lgreen}{RGB}{46, 139, 87} 
\usepackage{ulem}

\newcommand{\bs}[1]{\boldsymbol{#1}} 
\newcommand{\tr}[1]{\mathrm{Tr}\left[#1\right]}

\newcommand{\ut}[1]{\undertilde{#1}} 
\newcommand{\ve}[0]{\varepsilon} 
\newcommand{\vphi}[0]{\varphi} 
\newcommand{\thsm}[0]{\theta_{\mathrm{SM}}} 

\newcommand{\ua}[0]{\uparrow} \newcommand{\da}[0]{\downarrow}

\begin{document}


\title{Anomalous inverse proximity effect 
in unconventional-superconductor junctions}


\author{Shu-Ichiro Suzuki$^1$}
\author{Takashi Hirai$^2$}%
\author{Matthias Eschrig$^3$}%
\author{Yukio Tanaka$^1$}
\affiliation{$^1$Department of Applied Physics, Nagoya University, Nagoya 464-8603, Japan }
\affiliation{$^2$Independent researcher, Gifu, Japan}
\affiliation{$^3$Institute of Physics, University of Greifswald,
Felix-Hausdorff-Strasse 6, 17489 Greifswald, Germany}

\date{\today}

\begin{abstract}
We investigate the effects of Andreev bound states due to the unconventional pairing on the inverse proximity effect of ferromagnet/superconductor junctions. 
Utilizing quasiclassical Eilenberger theory, we
obtain the magnetization penetrating into the superconductor. 
We show that in a wide parameter range the direction of the induced magnetization is 
determined by two factors: whether Andreev bound states are present at the junction interface and the sign of the spin-mixing angle. 
In particular, when Andreev bound states appear at the interface, the direction of the
induced magnetization is opposite to that without Andreev bound states. 
We also clarify the conditions under which the inverted induced magnetization appears.
Applying this novel effect helps distinguishing the pairing symmetry of a superconductor. 
\end{abstract}

\pacs{Valid PACS appear here}
\maketitle

\section{Introduction}


Superconductors (SCs) dominated by an exotic pairing interaction are
often so-called unconventional superconductors (USCs), which break apart from the global phase symmetry one or more additional
symmetries of the normal state. In a conventional SC, the electrons
usually form Cooper pairs due to the retarded attractive effective
interaction resulting from electron-phonon coupling \cite{BCS}. On the
other hand, a repulsive interaction like the Coulomb interaction in
strongly correlated superconductors requires the order parameter to
change sign on the Fermi surfaces, resulting in anisotropic pairings
like for example the $d$-wave pairing in high-$T_c$ cuprate SCs and in
heavy-fermion SCs. 

The internal phase of the pair potential plays an important role in
forming  Andreev bound states (ABSs) \cite{ABS, Hara_PTP_1986,
Fu_PRL_94}.  At an interface of an USC, an ABS can be formed by the
interference between  incoming and outgoing quasiparticles where the
two quasiparticles feel different pair potentials depending on the
direction of motion \cite{Tanaka_PRL_95, Kashiwaya_review_00,
L_fwander_2001}.
Emergence of interface ABSs changes the properties of
superconducting junctions such as the transport properties
\cite{
Tanaka_PRL_95, TanaKashi_PRB_1997, Asano_PRB_2001, Eschrig_PRL_2003, Tanaka_PRB_2004, Asano_PRB_2004, Kwon_2004, Krawiec_PRB_04, Kopu_PRB_04,
Eschrig_ASSP_04, Buzdin_RMP_05, Bergeret_RMP_05, Asano_PRB_2006,
Asano_PRL_2007_2, Kaizer_Nature_2006, Eschrig_Nat_2007,
Asano_PRL_2007, Braude_Nazarov_PRL_2007, Houzet_PRB_07, Zaikin_PRB_08,
Zhao_PRB_08, Zaikin_PRB_10, Lofwander_PRL_10, Metalidis_PRB_10,
eschrig_2011_spin, Tanaka_review_2012, Machon_PRL_13, Machon_NJP_14,
Ozaeta_PRL_14, Bernardo_NatCom_15, Linder_NatPhys_15, Eschrig_2015, Eschrig_Royal_18, 
SIS_PRB_20} and magnetic
response \cite{Higashitani_JPSJ_97, Tanaka_PRBR_2005, Asano_PRL_2011,
Yokoyama_PRL_2011, SIS1, Bernardo_PRX_15, SIS2, Asano_PRB_15,   SIS3,
ShuPingLee_PRB_2017, Krieger_PRL_20, Higashitani_PRL_2013,
Higashitani_PRB_14}. 

The ABSs can drastically change the proximity effect as well. The
proximity effect is the penetration of Cooper pairs into the normal
metal (N) attached to a SC \cite{Deutscher_deGennes}. The conventional
proximity effect makes the density of states (DOS) in the N gapped
\cite{Volkov_Physica_1993, Golubov_JLTP_1988, Golubov_JETP_1989,
Golubov_PRB_1997, SIS19} (i.e., minigap), whereas a zero-energy peak
(ZEP) appears in the DOS when ABSs appear \cite{Tanaka_PRB_2004, Tanaka_PRBR_2005, Tanaka_PRB_2007, SIS19}.
Together with ABSs, odd-frequency Cooper pairs \cite{Berezinskii,
Balatsky_PRB_1992, Bergeret_RMP_05} are known to be induced
simultaneously. Odd-frequency pairs are demonstrated to show anomalous
response to the vector \cite{
Tanaka_PRBR_2005, 
Asano_PRL_2011, Yokoyama_PRL_2011, SIS1, Bernardo_PRX_15, SIS2,
Asano_PRB_15, Fominov_PRB_2015, SIS3, ShuPingLee_PRB_2017, Krieger_PRL_20} and
Zeeman potentials \cite{Higashitani_PRL_2013, Higashitani_PRB_14}. 

When an SC is in contact with  a magnetic material, another type of
proximity effect occurs. In a ferromagnet/SC (F/SC) junction, the
magnetization in the F penetrates into the SC on the length scale of
$\xi_0$ with $\xi_0$ being the superconducting coherence length. This
effect is called the inverse proximity effect (IPE)
\cite{Tokuyasu_PRB_88, Bergeret_PRB_04, Tollis_PRB_05,
Bergeret_PRB_05, Salikhov_PRL_2009, Xia_PRL_2009, Eschrig_PRB_13,Yagovstsev_2020}. 
The IPE has been
studied for junctions of  conventional SCs. The IPE was first studied
in a ballistic junction of a ferromagnetic insulator (FI) and an
$s$-wave SC \cite{Tokuyasu_PRB_88}. In this case, the induced
magnetization is antiparallel to the magnetization in the FI. If we
employ a ferromagnetic metal (FM) instead of a FI, the induced
magnetization is antiparallel in the diffusive limit
\cite{Bergeret_PRB_04}, whereas it is parallel in the ballistic limit
\cite{Tollis_PRB_05, Bergeret_PRB_05}.  
The IPE in conventional SC structures has been observed by several
experimental techniques, e.g. ferromagnetic resonance \cite{FMR_1998,FMR_2002},
nuclear magnetic resonance \cite{Salikhov_PRL_2009}, and polar
Kerr effect \cite{Xia_PRL_2009}. 
How anisotropic pairing in USCs affects the IPE, in contrast, has not been discussed so far.
In particular, induced ABSs and corresponding odd-frequency pairs are expected to affect how the magnetization penetrates into the SC.

\begin{figure}[b]
    \centering
		\includegraphics[width=0.48\textwidth]{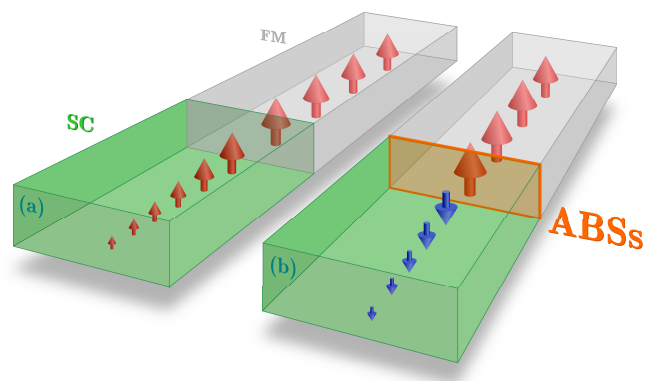}
		\caption{Schematics of the inverse proximity effect (IPE). The arrows
		represent the local magnetization vectors.  The ABSs are absent
		and present at the interface in (a) and (b), respectively.
        When there are ABSs, the magnetization induced in the superconductor (SC) is opposite to that without ABS. The length scale
		of the IPE is characterized by the superconducting coherence
		length $\xi_0$. 
		}
	\label{fig:sche}
\end{figure}

In this paper, we theoretically study the IPE in
F/USC junctions utilizing the quasiclassical Green's function theory.
We show that, when ABSs appear, the IPE induces a local magnetization
with the opposite sign compared to that in the F/conventional-SC
junction (see Fig.~\ref{fig:sche}).  Using a 1D model, we show that
the direction of the induced magnetization is determined by  two
factors: existence of ABSs and the sign of the spin-mixing angle. We
also show that spin-singlet and spin-triplet pairs near the interface
show a correspondence when $T \sim T_c$. Odd-frequency spin-triplet
$s$-wave pairs induced in a spin-triplet $p$-wave junction behave in
the same way as  spin-singlet $s$-wave pairs induced  in an $s$-wave
junction. 

In a two-dimensional system, the sign of the spin-mixing angle in general
depends on the momentum parallel to the interface, $k_\parallel$. When
the magnetization in  F is sufficiently small, the results are
qualitatively the same as those in  1D system. In the case of a large  magnetization in F, however, the $k_\parallel$ dependence of the spin-mixing angle
cannot usually be ignored. As a result, in this latter case, the direction of the induced magnetization for superconductors with nodes at $k_\parallel=0$ is
not simply determined by the two factors discussed in the 1D limit.


\section{Model and Formulation}

In this paper, we consider two-dimensional ballistic superconducting
junctions as shown in Fig.~\ref{fig:sche}, where the interface is located at $x=0$. 
We discuss the magnetization induced at the interface of an F/SC junction, where F and SC stand for  ferromagnet and  superconductor. 

Superconductivity in the ballistic limit can be described by the
quasiclassical Eilenberger theory. The Green functions obey the Eilenberger equation: 
\begin{align}
  & i \bs{v}_F \cdot \bs{\nabla} \check{g}
	+ \left[ \check{H},~\check{g} \right]_-
	= 0, 
	\label{eq:Eilen}
	\\[2mm]
	& \check {g} = \left( \begin{array}{rr}
	 \hat{g} &  
	 \hat{f} \\[1mm]
	-\hat{\ut{f}} & 
	-\hat{\ut{g}} \\
	\end{array} \right), 
	\hspace{4mm}
	%
  \check {H} = \left( \begin{array}{cc}
	i\omega_n\hat{\sigma}_0 &  
	\hat{\eta} \\[1mm]
	\hat{\ut{\eta}} & 
	-i\omega_n\hat{\sigma}_0 
	\end{array} \right), 
\end{align}
where $\check{g}=\check {g}(x,\bs{k},i\omega_n)$ is the Matsubara 
Green function and $\bs{v}_F$ is the Fermi velocity. In this paper, the accents $\check{\;}$
and $\hat{\;}$ denote matrices in particle-hole space and spin
space, respectively.  The identity matrices in particle-hole and spin space are denoted by $\check{\tau}_0$ and $\hat{\sigma}_0$.  The Pauli matrices in particle-hole space and in spin space are $\check{\tau}_j$ and $\hat{\sigma}_j$ with $j \in \{1,2,3\}$, respectively. 
All of the functions satisfy the symmetry relation 
$\hat{    K } (x, \bs{k},i\omega_n) =
[\hat{\ut{K}} (x,-\bs{k},i\omega_n)]^*$, where the unit vector
$\bs{k}$ represents the direction of the Fermi momentum. 
The effects of the F can be taken into account through  boundary
conditions. 


The Eilenberger equation \eqref{eq:Eilen} must be supplemented by a normalization condition, $\check g^2=\check 1$, which is nonlinear. It can be implemented explicitely by the
so-called Riccati parametrization \cite{Schopohl_PRB_95, Eschrig_PRB_99, 
Eschrig_PRB_00, Eschrig_PRB_09, Eschrig_NJP_15}. The Green function can be
expressed in terms of the coherence function $\hat{\gamma}$ in the following way \cite{Eschrig_PRB_09, Eschrig_NJP_15}:
\begin{align}
	& \check {g} = 2 \left( \begin{array}{rr}
	 \hat{\mathcal{G}} &  
	 \hat{\mathcal{F}} \\[1mm]
	-\hat{\ut{\mathcal{F}}} & 
	-\hat{\ut{\mathcal{G}}} \\
	\end{array} \right) - \check{\tau}_3, 
	\label{eq:Ric-Para0}
	%
	\end{align}
	\begin{align}
	& \hat{\mathcal{G}} = (1-\hat{\gamma} \hat{\ut{\gamma}} )^{-1},\hspace{6mm}
	  \hat{\mathcal{F}} = (1-\hat{\gamma} \hat{\ut{\gamma}} )^{-1}\hat{\gamma},\\
	& \hat{\ut{\mathcal{G}}} = (1- \hat{\ut{\gamma}}\hat{\gamma} )^{-1},\hspace{6mm}
	  \hat{\ut{\mathcal{F}}} = (1- \hat{\ut{\gamma}}\hat{\gamma}
		)^{-1}\hat{\ut{\gamma}},
\label{eq:Ric-Para}
\end{align}
where $\hat{\gamma}= \hat{\gamma}(x, \bs{k}, i \omega_n)$. 
The Riccati parametrization reduces the
Eilenberger equation \eqref{eq:Eilen} into the Riccati-type
differential equation \cite{Eschrig_PRB_99}: 
\begin{align}
  & i \bs{v}_F \cdot \bs{\nabla} \hat{\gamma}
	+ \hat{\xi} \hat{\gamma} - \hat{\gamma}\hat{\ut{\xi}}
	-\hat{{\eta}}+ \hat{\gamma} \hat{\ut{\eta}} \hat{\gamma}= 0. 
	\label{eq:Riccati01}
\end{align}
This Riccati-Eilenberger equation can be simplified as  
\begin{align}
  & \bs{v}_F \cdot \bs{\nabla} \gamma
	+ 2 \omega_n \gamma
	- \Delta_{\bs{k}}
	+ \Delta_{\bs{k}}^* \gamma^2= 0, 
	\label{eq:Riccati01-1}
\end{align}
where we assume the superconductor has only one spin component [i.e., 
$\hat{\eta} = i \Delta_{\bs{k}}
(i \hat{\sigma}_\nu \hat{\sigma}_2)$ with $\nu \in \{0,1,2,3\}$]
and the spin structure of the coherence function is parameterized as 
$\hat{\gamma} = \gamma (i \hat{\sigma}_\nu \hat{\sigma}_2)$. 
In a homogeneous superconductor, the coherence function is given by 
\begin{align}
  \bar{\gamma}(\bs{k}, i \omega_n)
	= \frac{\Delta_{\bs{k}}}{\omega_n +
	\sqrt{\omega_n^2+|\Delta_{\bs{k}}|^2}}, 
	\label{eq:gumma_ene_bul}
	%
\end{align}
where the coherence function needs to satisfy the condition $
\lim_{\omega_n \to \infty} \gamma = 0$. 

In USCs, the pair potential depends on the direction of the momentum. 
For isotropic Fermi surfaces, the momentum dependence of the pair potential is given by 
\begin{align}
  \Delta_{\bs{k}} = \left\{
	\begin{array}{ll}
	\Delta_0               & ~\text{for           $s$ wave,}\\
	\Delta_0 \cos \vphi    & ~\text{for         $p_x$ wave,}\\
	\Delta_0 \sin \vphi    & ~\text{for         $p_y$ wave,}\\
	\Delta_0 \cos (2\vphi) & ~\text{for $d_{x^2-y^2}$ wave,}\\
	\Delta_0 \sin (2\vphi) & ~\text{for      $d_{xy}$ wave,}\\
	\end{array}
	\right.
\label{}
\end{align}
where $\vphi$ characterizes the direction of the Fermi momentum; $k_x =
\cos \vphi$ and $k_y = \sin \vphi$. The temperature
dependence of the pair potential is determined by the self-consistency
condition for a homogeneous SC: 
\begin{align}
  & \Delta_0(T)
	= 2 N_0 \lambda 
	\frac{\pi}{\beta}
	\sum_{\omega_n>0}^{\omega_c}
	\int
  \frac{\Delta_\vphi \Lambda_\vphi}{\sqrt{\omega_n^2+\Delta_\vphi^2}}
	\frac{d \vphi}{2 \pi}
	\label{gap}
\end{align}
where $N_0$ is the density of states (DOS) per spin
at the Fermi level, $\beta=1/T$,  $\omega_c$ is the BCS cutoff energy,
and 
\begin{align}
  \Lambda_{\vphi} = \left\{
	\begin{array}{ll}
	  1             & ~\text{for           $s$ wave,}\\
	  2\cos \vphi    & ~\text{for         $p_x$ wave,}\\
	  2\sin \vphi    & ~\text{for         $p_y$ wave,}\\
	  2\cos (2\vphi) & ~\text{for $d_{x^2-y^2}$ wave,}\\
	  2\sin (2\vphi) & ~\text{for      $d_{xy}$ wave.}\\
	\end{array}
	\right.
\end{align}
The coupling constant $\lambda$ is given by
\begin{align}
	\lambda
	=  
	\frac{1}{N_0 }
  \left[ \mathrm{ln}
  \left( \frac{T}{T_c} \right)
  +\sum_{n = 0}^{n_c}
	\frac{1}{n+1/2}
	\right]^{-1}, 
	\label{couple}
\end{align}
with  $ n_c = \omega_c / 2 \pi T $.

In this paper, the temperature dependence of the pair potential is taken into account, however the approximation of a spatially homogeneous pair potential is made.

Magnetic potentials can induce the local magnetization in the 
SC. The induced magnetization is defined as 
\begin{align}
  & M(x) = \mu_B( n_\ua - n_\da), 
	\\
  & n_\alpha(x) = \langle 
  \Psi^\dagger_\alpha(x)
  \Psi        _\alpha(x)
	\rangle
\end{align}
where $\mu_B$ is the effective Bohr magneton, $\alpha=$~$\ua$ or $\da$ is the spin
index, 
and $\Psi_\alpha$ ($\Psi^\dagger_\alpha$) is the annihilation
(creation) operator of a particle with the spin $\alpha$. This local
magnetization can be obtained from the diagonal parts of the quasiclassical
Green's function: 
\begin{align}
  M(x)
	&= \frac{\mu_B N_0 \pi}{2i\beta}
  \sum_{\omega_n} \int \tr{ 
	(\tau_3 \otimes {\sigma}_3 ) \hat{g}
	(x, \vphi, i\omega_n) } \frac{d \vphi}{2\pi} , 
  \label{eq:mag5}
	\nonumber \\
	&= 2 \pi \mu_B N_0 T 
  \sum_{n=0}^{n_c} \int 
	 \mathrm{Im}[g_\ua - g_\da]
	 \frac{d \vphi}{2\pi}, 
\end{align}
where we have used the symmetry of the Matsubara Green function 
$\hat{g}(x, \vphi, i\omega_n) = -{\hat{g}}^*(x, \vphi,-i\omega_n)$, and the abbreviation
$\hat{g} = \mathrm{diag}[g_+, g_-]$ with $g_\pm$ being the normal
Green's function for the up and down spin. 

\begin{figure}[tb]
  \centering
  \includegraphics[width=0.48\textwidth]{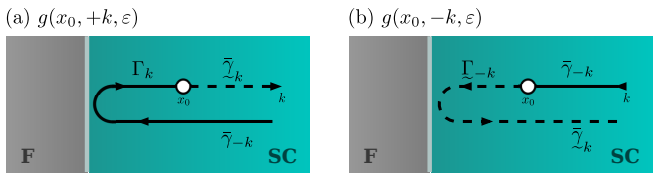}
	\caption{Quasiclassical path to obtain the Green's function in
	$(+k,k_y)$ and $(-k,k_y)$ direction, respectively. The particle-like
	(hole-like) coherence function must be solved along the
	quasiclassical path in $+k$ ($-k$) direction, which are indicated by
	the solid and broken lines. The momentum parallel to the interface
	is conserved during the reflection. The overbar symbol $\bar{\cdot}$ denotes the bulk
	value of the coherence function. }
  \label{fig:path}
\end{figure}

\subsection{Boundary condition}
The boundary conditions for the coherence functions are given in
Refs.~\cite{ 
Eschrig_PRB_00,
Eschrig_PRB_09,
Eschrig_NJP_15}. 
Hereafter, the outgoing (incoming) coherence functions are denoted by 
$\Gamma$ ($\gamma$) as introduced in Ref.~\cite{Eschrig_PRB_00}.
When the SC and F are semi-infinitely long in the $x$ direction, the
boundary condition is simplified because $\gamma = 0$ in the F. 
The boundary condition is given by 
\begin{align}
  \hat{\Gamma} = 
	\hat{r}
	\hat{\gamma} 
	\hat{r}^*, 
  \label{eq:bc02} 
\end{align}
where 
the reflection-coefficient matrix $\hat{r}$ is given by 
\begin{align}
  \hat{r} =
	\left[ \begin{array}{cc}
	  r_{\ua} & 
	   \\
	   & 
	  r_{\da} 
	\end{array} \right]
	=
	\left[ \begin{array}{cc}
	  |r_{\ua}| e^{i(\phi+\theta_{\mathrm{SM}})}& 
	   \\
	   & 
	  |r_{\da}| e^{i(\phi-\theta_{\mathrm{SM}})}
	\end{array} \right]. 
\end{align}
The angle $\thsm$ is the so-called spin-mixing angle \cite{Tokuyasu_PRB_88} and 
$r_{\ua}$ and $r_{\da}$ are the
reflection coefficients for the up-spin and down-spin particles injected
from the SC side. 
The reflection coefficients can be obtained by matching
the wave functions: 
\begin{align}
	r_{\alpha} & = 
	\frac{\hbar(v-v_\alpha)-2iV}
	     {\hbar(v+v_\alpha)+2iV},
			 \label{eq:reflec}
\end{align}
where we have assumed the potential barrier $V\delta(x)$ at the interface 
and $v=\hbar k /m$ ($v_\alpha = \hbar k_\alpha /m$) is the $x$ component of the Fermi velocity 
in the SC (F) side; 
$k = (2\mu_S-k_y^2)^{1/2}$ and 
$k_{\ua(\da)} = \{ 2m [\mu_F +(-) M_0] -k_y^2 \}^{1/2}$ with $m$ being 
the effective mass of a quasiparticle. Note that we have made $\hbar $ explicit 
to avoid misunderstandings.

In a single-spin-component superconductor, the coherence amplitude
in the bulk can be expressed as 
\begin{align}
      \hat{\gamma}  =     \gamma (i\hat{\sigma}_\nu \hat{\sigma}_2), \hspace{8mm}
  \ut{\hat{\gamma}} = \ut{\gamma}(i\hat{\sigma}_\nu \hat{\sigma}_2)^\dagger. 
\end{align}
In the present case (i.e., magnetic potentials are parallel to the
spin quantization axis), the reflection coefficients are diagonal in
spin space (i.e., $\hat{r} = \mathrm{diag}[r_{\ua}, r_{\da}]$). 
Therefore, the outgoing coherence functions \eqref{eq:bc02} are given by 
\begin{align}
	& \hat{\Gamma}
	= \left( \begin{array}{rr}
	   & 
        \Gamma_{\ua} \\
  s_\nu \Gamma_{\da} &
	   \\
	\end{array} \right)
	=
	\left( \begin{array}{rr}
	   & 
        r_{\ua } \gamma r^*_{\da} \\
  s_\nu r_{\da } \gamma r^*_{\ua} &
	   \\
	\end{array} \right), 
  \label{eq:bc03} 
\end{align}
for the opposite-spin pairing ($\nu = 0$ or $3$), where $s_\nu = 1$ ($-1$) for the
spin-triplet (singlet) pairing. 
The boundary conditions 
obtained here are consistent with those derived using the so-called evolution
operators \cite{AAHN_PRB_89, Nagato_JLTP_93, Tanuma_PRB_01, Hirai_PRB_03}.


To obtain the coherence amplitude, we need to consider the group velocity
of the quasiparticle and quasihole properly. 
The quasiclassical paths
to obtain $\check{g}(x_0, k, k_y, \ve)$ and  $\check{g}(x_0, -k, k_y, \ve)$ are
shown in Figs.~\ref{fig:path}(a) and \ref{fig:path}(b), where the solid and
broken lines represent the path for the particle-like and hole-like
coherence amplitudes and the direction of the arrows indicate the
direction of the Fermi momentum. Since the quasiparticle (quasihole)
propagates in
the same (opposite) directions as $\bs{k}$, $\hat{\gamma}$
and $\ut{\hat{\gamma}}$ should be solved in $\bs{k}$ and $-\bs{k}$
direction respectively.

The Green's function can be obtained from the coherence functions [see Eq.~\eqref{eq:Ric-Para0}]. 
Using the boundary condition, the diagonal part of the Green's
functions at the interface are given in terms of the coherence functions:
\begin{align}
  & \hat{g}_{+k} = (1-\hat{\Gamma} \hat{\ut{\gamma}})^{-1} (1+\hat{\Gamma} \hat{\ut{\gamma}}), \\
  & \hat{g}_{-k} = (1-\hat{\gamma} \hat{\ut{\Gamma}})^{-1} (1+\hat{\gamma} \hat{\ut{\Gamma}}), 
\end{align}
where $\hat{g}_{\pm k}  = \hat{g}( x=0^+, \pm k, k_y, i \omega_n)$. 
The spin-reduced Green's functions at the interface are 
\begin{align}
  & {g}_{\alpha,{+k}}
	= \frac{ 1+\Gamma_\alpha {\ut{\gamma}}}{1-\Gamma_\alpha {\ut{\gamma}}}, 
	\hspace{2mm}
    {g}_{\alpha,{-k}}
	= \frac{ 1+{\gamma} \ut{\Gamma}_{\alpha'} }{ 1-{\gamma}
	\ut{\Gamma}_{\alpha'} }, 
	\label{eq:sr}
\end{align}
where $\alpha'$ means the
opposite spin of $\alpha$. The spin structure of the
coherence functions are parameterized as $\hat{\Gamma} = \mathrm{diag}[\Gamma_\ua,
\Gamma_\da](i \hat{\sigma}_\nu \hat{\sigma}_2)$ and 
$\hat{\ut{\Gamma}} = \mathrm{diag}[\ut{\Gamma}_\ua,
\ut{\Gamma}_\da](i \hat{\sigma}_\nu \hat{\sigma}_2)^\dagger$. 
Assuming the spatially homogeneous pair potential, we can replace
$\gamma$ in Eq.~\eqref{eq:sr} by $\bar{\gamma}$, where the symbol $\bar{\;}$ means bulk values. 
This assumption changes the results only quantitatively but not
qualitatively. 

%

\begin{figure}[tb]
  \vspace{0mm}
  \centering
  \includegraphics[width=0.46\textwidth]{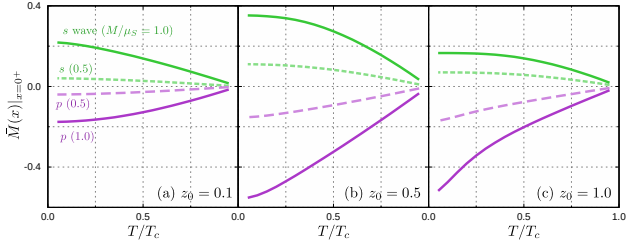}
	\caption{Temperature dependence of induced magnetization at the
	interface of a \textit{1D} FM/SC junction. The induced magnetization
	is normalized as $\bar{M} = M / 2\pi \mu_B N_0 T_c$. 
	The magnetization of the F is set to $M_0=\mu_F$ or $0.5\mu_F$ with
	$\mu_F = \mu_S$. 
	The spin-independent barrier potential is set to (a) $z_0 = 0.1$,
	(b) 0.5, and (c) 1.0, where $\theta_{\mathrm{SM}}<0$ for
	all of the sets of the parameters. 
	The induced magnetization	of the $s$-wave junctions is positive,
	whereas that of the $p$-wave junctions are negative. 
	The pair potential depends on the temperature
	 but is kept constant as function of the spatial coordinate $x$. }
  \label{fig:magFM1D}
\end{figure}
\begin{figure}[t]
  \vspace{0mm}
  \centering
  \includegraphics[width=0.36\textwidth]{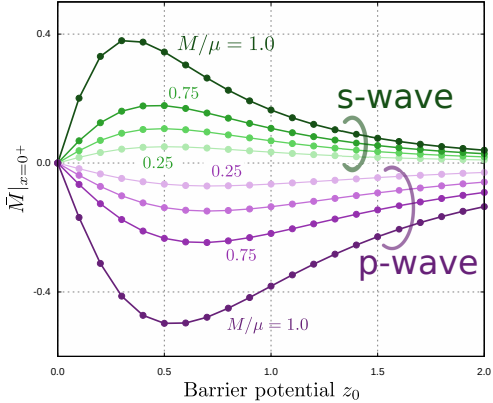}
	\caption{Induced magnetization at the interface of a \textit{1D} FM/SC
	junction. The magnetization $ M |_{x=0^+}$
	for $s$- and $p$-wave superconductors are respectively positive and negative 
	regardless of the magnitude of $z_0$. 
	The temperature is set to $T=0.2T_c$. }
  \label{fig:magFM1DZ}
\end{figure}

\begin{figure}[tb]
  \vspace{0mm}
  \centering
  \includegraphics[width=0.48\textwidth]{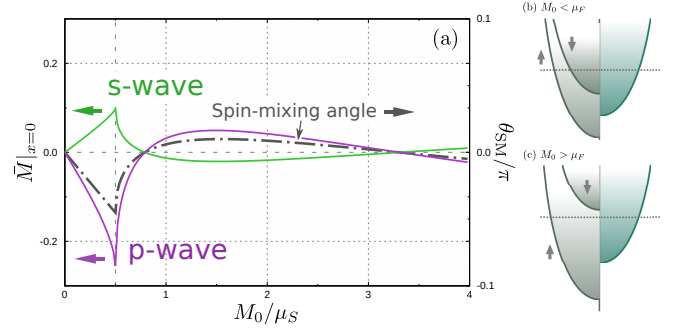}
	\caption{Induced magnetization and spin-mixing angle 
	of a \textit{1D} F/SC
	junction as a function of $M_0$. 
	The chemical potential in the F is set to $\mu_F = 0.5\mu_S$, which
	means that the F is an FM (HM) when $M_0 < 0.5 \mu_S$ ($M_0 \geq 0.5\mu_S$). 
	The schematic band structures of the FM/SC and HM/SC
	junctions are shown in (b) and (c), respectively. 
	The sign of the magnetization	is determined by $\mathrm{sgn}(M_0)$ and
	the pairing symmetry. 
	The temperature and the barrier potentials are set to $T=0.2T_c$ and $z_0=1$. }
  \label{fig:magFMM}
\end{figure}
\begin{figure}[tb]
  \vspace{0mm}
  \centering
  \includegraphics[width=0.48\textwidth]{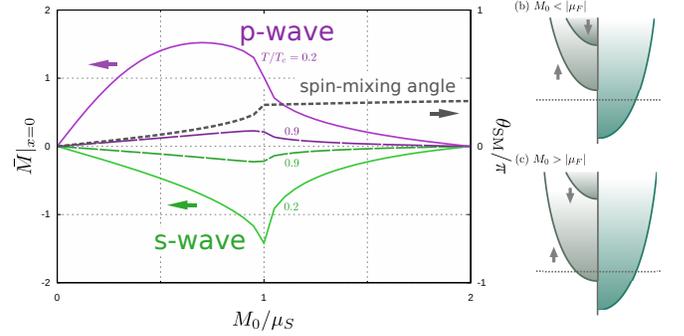}
	\caption{Induced magnetization and spin-mixing angle of a
	\textit{1D} F/SC junction with $\mu_F = -\mu_S$.  The F is
	insulating for $M_0 < \mu_S$ or a half-metallic for $M_0 > \mu_S$. 
	The induced magnetization	for an $s$-wave junction is negative 
	regardless of the temperature (i.e., $T=0.2T_c$ and $0.9T_c$),
	whereas that $p$-wave junctions are positive. 
	The barrier potential is set to $z_0=0$. 
	The $s$-wave results are consistent with those in 
	Ref.~\onlinecite{Tokuyasu_PRB_88}, where $\theta_{\mathrm{SM}}$ is
	defined with the opposite sign compared with our definition. 
	The schematics of band structures of the FI/SC and HM/SC
	junctions are shown in (b) and (c), respectively. }
  \label{fig:magFI}
\end{figure}
\section{One-dimensional system}

In order to understand the basics of the IPE. 
We start with one-dimensional (1D) systems. Such systems can be
considered by setting $k_y=0$. 
\subsection{Ferromagnetic-metal junction}

The temperature dependence of the induced magnetizations at the interface of the F/SC junction 
are shown in Fig.~\ref{fig:magFM1D} where spin-singlet even-parity
and spin-triplet odd-parity superconducting junctions are considered,
which correspond to the $s$- and $p_x$-wave superconducting junctions
in the 2D case respectively. The
magnetization in the F is set to $M_0=\mu_F$ or $0.5\mu_F$ with $\mu_F =
\mu_S$, the barrier potential $z_0 = mV/k$ is set to 
(a) $z_0=0.1$, 
(b) $z_0=0.5$, and 
(c) $z_0=1.0$, and the pair potential is assumed spatially homogeneous but
temperature dependent. 
In the even-parity case, the magnetization in the F induces the
\textit{parallel} magnetization in the SC as shown in
Fig.~\ref{fig:magFM1D}. This behavior is consistent
with that in the ballistic limit in Refs.~\cite{Tollis_PRB_05,
Bergeret_PRB_05}. In the odd-parity case, on the contrary, the induced
magnetization is \textit{antiparallel} to $M_0$. 
We have confirmed that no magnetization is induced when the
$\bs{d}$-vector is perpendicular to the magnetization vector in the F. 

The $z_0$ dependences of $M|_{x=0^+}$ are shown in Fig.~\ref{fig:magFM1DZ}. The
induced magnetization is not a monotonic function of the barrier
parameter $z_0$. 
When $z_0 \to \infty$, the reflection coefficient \eqref{eq:reflec}
becomes spin-independent (i.e., $r_\alpha \to -1$). Therefore,
$M_{x=0^+}$ vanishes in this limit. 
When $z_0 = 0$, 
the reflection coefficients in Eq.~\eqref{eq:reflec} are real as long as $k_\alpha$ is real
(i.e., F is a ferromagnetic metal), which means that the 
reflected quasiparticles do not have an additional spin-dependent phase shift. 
As a result, no magnetization is induced in the SC \footnote{Note that we have confirmed $\thsm < 0$ for all set of $z_0$ and $M_0$}. 

The induced magnetization $M|_{x=0^+}$ and the spin-mixing angle $\thsm$ as functions of
$M_0$ are shown in
Fig.~\ref{fig:magFMM}(a) where $\mu_F = 0.5\mu_S$, $z_0 = 1.0$, and $T=0.2T_c$.
In this case, the F is 
a ferromagnetic metal (FM) for $0<M_0<\mu_F$ and 
a half-metal (HM) for $M_0>\mu_F$ as schematically illustrated in 
Figs.~\ref{fig:magFMM}(b) and \ref{fig:magFMM}(c). As shown in 
Fig.~\ref{fig:magFMM}(a), the sign of $M|_{x=0^+}$ 
for the $s$-wave junction is always opposite to $\mathrm{sgn}[\thsm]$, 
whereas that for the $p$-wave junction always has the same sign. 
These results demonstrate that the sign of the induced magnetization
is determined by the two factors: the pairing symmetry and the sign of
the spin-mixing angle $\thsm$.


\begin{figure}[tb]
  \vspace{0mm}
  \centering
  \includegraphics[width=0.48\textwidth]{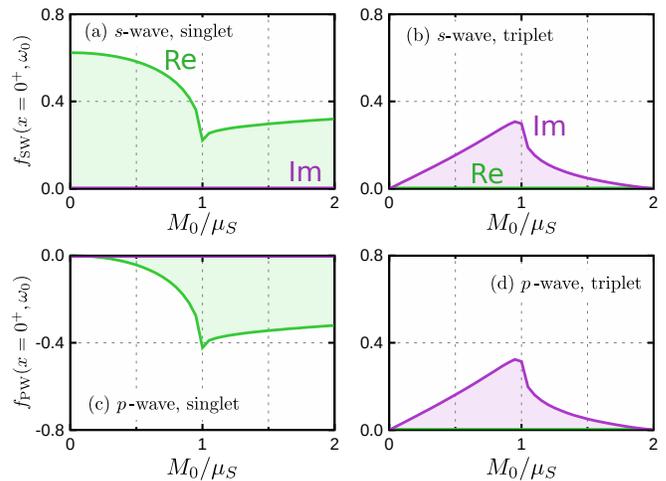}
	\caption{Pair amplitudes in a \textit{1D} F/$s$-wave junction. The
	$s$-wave spin-singlet, $s$-wave triplet, $p$\,-wave singlet, and
	$p$\,-wave triplet components are plotted in (a), (b), (c), and (d). 
	When $M_0 < \mu_S$ (i.e., FI regime), 
	the magnetization is mainly generated by the product of the $s$-wave
	singlet and $s$-wave triplet pairs because the $s$-wave singlet is
	dominant [see Eq.~\eqref{eq:pair}]. 
	%
	The temperature and the Matsubara frequency are set to $T=0.9T_c$
	and $\omega_n = \omega_0$. }
  \label{fig:FImag_pair_s}
\end{figure}

\subsection{Ferromagnetic-insulator junction}
The IPE occurs in 
ferromagnetic-insulator(FI)/SC junctions as well.  The $M_0$
dependence of $M |_{x=0^+}$ is shown in Fig.~\ref{fig:magFI}(a). 
In order to model the FI, we set $\mu_F = -\mu_S$ where the FI-HM
transition occurs at $M_0 = |\mu_F|$ as schematically shown 
in Figs.~\ref{fig:magFI}(b) and \ref{fig:magFI}(c). 
Figure~\ref{fig:magFI}(a) shows that $M |_{x=0^+}$ in the $s$-wave junction is
antiparallel to $M_0$ in the F. This result is consistent with 
Ref.~\onlinecite{Tokuyasu_PRB_88} \footnote{The spin-mixing angle in
Ref.~\onlinecite{Tokuyasu_PRB_88} is defined with an extra minus sign
compared with ours.}.  For the $p$-wave case, on the other hand, the
induced magnetization is parallel to $M_0$. We can conclude that the
IPE induces the local magnetization with the
opposite sign compared with an $s$-wave junction. 

\begin{figure}[tb]
  \vspace{0mm}
  \centering
  \includegraphics[width=0.48\textwidth]{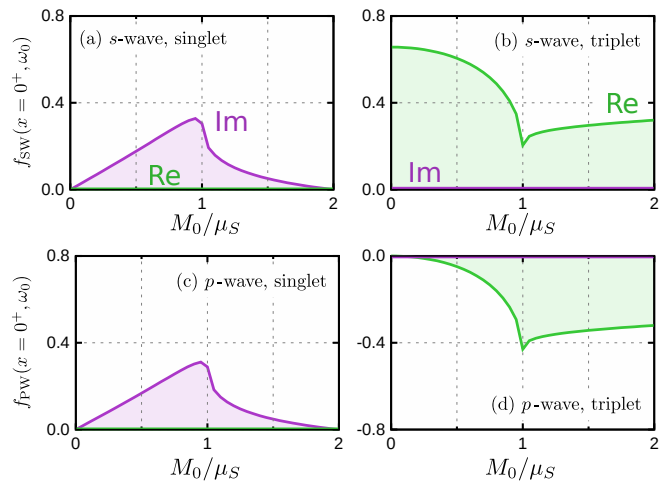}
	\caption{Pair amplitudes in a 1D 
	F/$p$\,-wave junction. The results are plotted in the same manner as
	in Fig.~\ref{fig:FImag_pair_s}. 
	The main contribution comes from the $s$-wave pairs even in the
	$p$\,-wave spin-triplet superconducting junction. }
  \label{fig:FImag_pair_p}
\end{figure}

 At low temperature, 
$|M|$ in $M_0<\mu_F$ for the $p$-wave junction is greatly larger than
that for the $s$-wave case. Similar low-temperature anomalies of
the magnetic response have been reported so far\cite{Asano_PRL_2011,
Yokoyama_PRL_2011, SIS1, SIS2, Asano_PRB_15, 
Higashitani_PRL_2013, SIS1, SIS3, Higashitani_PRB_14}. 
These anomalies are explained by the emergence of the zero-energy
ABSs. 
In superconducting junctions, ABSs appear because of the interference 
between the quasiparticle propagating into an interface and reflected one.
Therefore, the effects of ABSs become larger as increasing
reflection probability $|R|$ \cite{Kashiwaya_review_00}. In other words, the anomalous IPE
becomes more prominent in an FI/SC junction than that in FM/SC
and HM/SC junctions. 

The amplitude of $M|_{x=0^+}$ changes suddenly 
at $M_0=|\mu_F|$ in accordance with the FI-HM transition.
After the FI-HM transition, the induced magnetization decreases with
increasing $M_0$ and vanishes at $M_0=2|\mu_F|$ regardless of the
pairing symmetry. When  $\mu_F \pm M_0 =\mu_S$, the dispersion
relation of either band in the F becomes identical to that in the SC,
with the consequence that the reflection probability for the corresponding spin
becomes zero. In this case, both $\Gamma_\ua$ and $\Gamma_\da$  are 
zero [see Eq.~\eqref{eq:bc03}], which means the IPE does not occur. 

The induced magnetization can be expressed in terms of the pair
amplitude (i.e., anomalous Green's function) when $T \sim T_c$ (see
Appendix~\ref{sec:appen1} for the details) \cite{Higashitani_PRL_2013, Higashitani_JPSJ_2014}. 
In the 1D limit, in particular, the magnetization is give by 
\begin{align}
  & M
	\approx
	4 \pi \mu_B N_0 T \sum_{\omega_n>0} m_{0,3}, 
		\label{eq:pair}
		\\
  & m_{\nu, \nu'} = 
	\mathrm{Im}\big[ 
	  f  _{\nu,\mathrm{SW}}
		f^*_{\nu',\mathrm{SW}} 
	+ f  _{\nu,\mathrm{PW}} 
	  f^*_{\nu',\mathrm{PW}} \big], 
		\label{eq:pair2}
\end{align}
where SW and PW stand for the $s$-wave and $p$-wave pairings 
respectively. The spin indices $\nu=0$ and 3 represent the spin-singlet and
spin-triplet pairs respectively. Note that $f_{0,PW}$ and $f_{3,SW}$
should be odd-functions of $\omega_n$ to satisfy the Pauli rule
\cite{Eschrig_PRL_2003,Eschrig_Nat_2007,Eschrig_JLTP_2007}. 
Equation \eqref{eq:pair} means that the
magnetization is given by the product of the spin-singlet and
spin-triplet pairs.

\begin{figure}[tb]
  \vspace{0mm}
  \centering
  \includegraphics[height=0.30\textwidth]{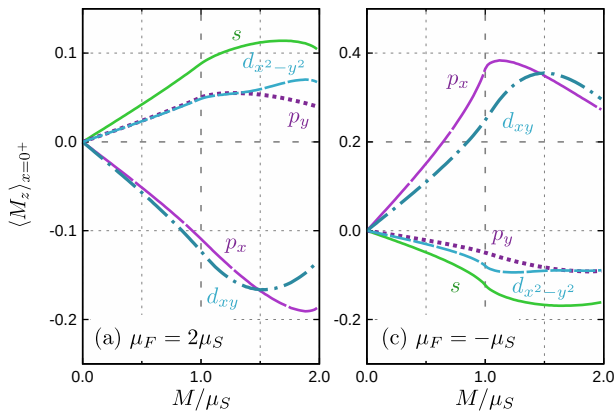}
	\caption{Induced magnetization of a \textit{2D} F/SC
	junction. 
	The chemical
	potential is set to (a) $\mu_F = 2\mu_S$ and (b) $\mu_F =
	-\mu_S$. 
	When the ABSs are present, the induced magnetizations have the
	opposite sign to those without ABSs. 
	The temperature and the barrier potentials are set to $T=0.2T_c$ and $z_0=1$. }
  \label{fig:FSmag}
\end{figure}
The pair amplitudes at the interface of the
$s$-wave junction are shown in Fig.~\ref{fig:FImag_pair_s}, where
$T=0.9T_c$ and $\omega = \omega_0 = \pi T$. 
The 
spin-singlet $s$-wave, 
spin-triplet $s$-wave, 
spin-singlet $p$-wave, and 
spin-triplet $p$-wave pair amplitudes 
are shown
in Figs.~\ref{fig:FImag_pair_s}(a), 
\ref{fig:FImag_pair_s}(b), 
\ref{fig:FImag_pair_s}(c), and \ref{fig:FImag_pair_s}(d). In
the FI region (i.e., $M_0 < \mu_F$) of an $s$-wave junction, the
conventional spin-singlet $s$-wave pairs are dominant and the other
pair amplitudes are relatively small. The magnetization in this case
is mainly generated by the spin-singlet and spin-triplet $s$-wave
Cooper pairs (i.e., $
	  f  _{0,\mathrm{SW}}
		f^*_{3,\mathrm{SW}} 
	\ll f  _{0,\mathrm{PW}} 
	  f^*_{3,\mathrm{PW}} 
$ when $M_0 \ll \mu_F$).

In the $p$-wave case, on the other hand, the spin-triplet $s$-wave
pair amplitude is dominant for $M_0<\mu_F$ as shown in
Fig.~\ref{fig:FImag_pair_p}.  In addition, the spin-triplet $s$-wave
pair amplitudes have almost the same $M_0$ dependences for the
spin-singlet $s$-wave pair amplitude as shown in 
Figs.~\ref{fig:FImag_pair_s}(a) and \ref{fig:FImag_pair_p}(b). 
Comparing
Figs.~\ref{fig:FImag_pair_s} and \ref{fig:FImag_pair_p}, similar
correspondences between spin-singlet and spin-triplet pairs are
confirmed. 
In Eq.~\eqref{eq:pair}, such a singlet-triplet conversion results in
the sign change of the magnetization (i.e., $m_{0,3} = - m_{3,0}$). 
Namely, 
the $s$-wave pairs in the $p$-wave junction generate almost the same
amplitude of the magnetization compared with the one in the $s$-wave junction.
The direction, however, is opposite compared with that in the $s$-wave junction. 
In the Cooper pair picture, 
the spin structure of the dominant Cooper pair determines the
direction of the induced magnetization. 
\begin{figure}[t]
  \vspace{0mm}
  \centering
  \includegraphics[height=0.30\textwidth]{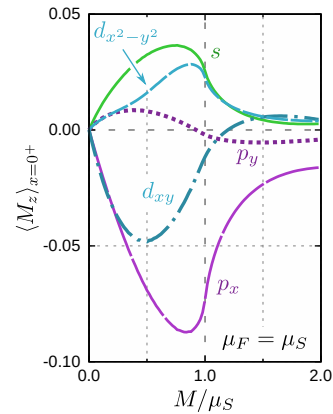}
	\caption{Induced magnetization of a 2D F/S
	junction with $\mu_F = \mu_S$. The results are plotted in the same
	manner in Fig.~\ref{fig:FSmag}. 
	When $M_0$ is sufficiently large, 
	the induced magnetizations for an $s$-wave and $p_y$-wave junctions
	are opposite even though no ABS appears in both of the junctions. 
	}
  \label{fig:FSmag2}
\end{figure}


\section{Two-dimensional system}

Most realistic superconducting junctions are two-dimensional or three-dimensional. In 2D systems, local quantities 
should be obtained via a $k_y$ integration, where $k_y$ is the
momentum parallel to the interface (see Eq.~\eqref{eq:mag5}). In particuar, the induced magnetization
generated by the IPE is obtained via $k_y$ integration where 
the partial magnetization depends on $k_y$ via the transport
coefficients and the pair potential. 

\begin{figure*}[t]
  \vspace{0mm}
  \centering
  \includegraphics[width=0.98\textwidth]{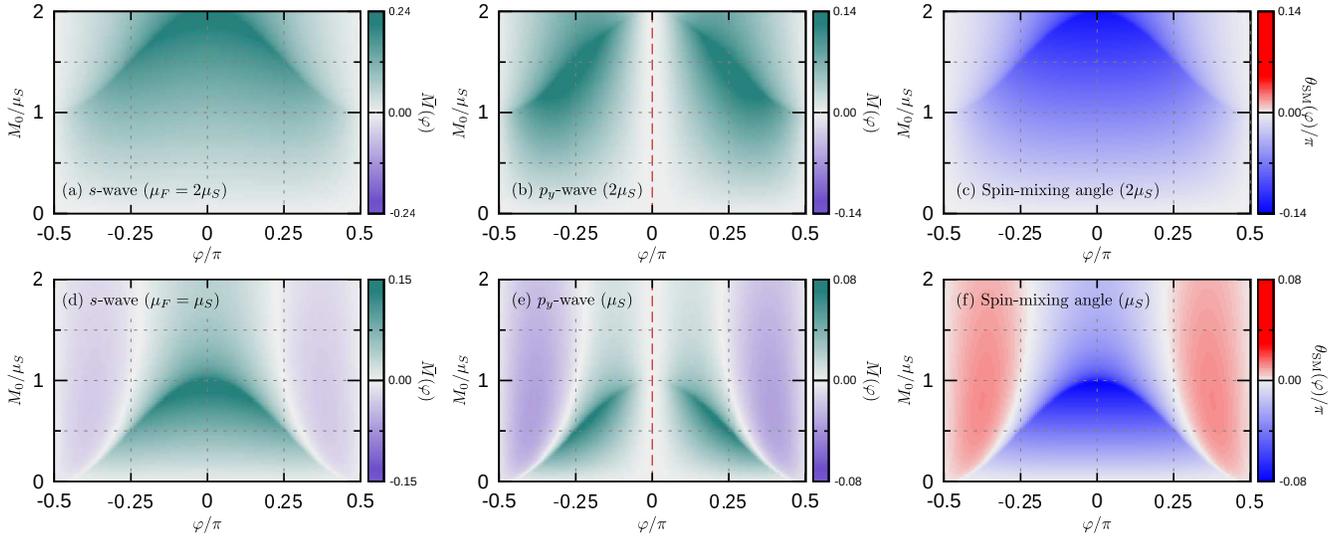}
	\caption{
	Angle-resolved magnetization and spin-mixing angle of  
	F/SC junctions. 
	The chemical potential in the F is set to 
	$\mu_F = 2\mu_S$ in (a)-(c), whereas $\mu_F = \mu_S$
    in (d)-(f).
	The order parameter is assumed spin-singlet $s$-wave in (a) and (d), and spin-triplet $p$-wave in (b) and (e). When $\mu_F = \mu_S$ and $M_0 \neq 0$, the sign change of $\thsm$ occurs around $k_y \sim \pm k_F$. In (e), when $M_0>\mu_F$, the positive contribution to $M$ is smaller than the negative one due to the nodes of the $p_y$-wave gap at $\vphi=0$. 
	In (a), (b), (d), and (e), the partial magnetizations are normalized to $\bar{M}(\vphi) = M(\vphi) / 2\pi \mu_B N_0 T_c$ and the magnetization in  F is changed from $M_0 = 0.8$ to $2.0$ by $0.2$. 
	The temperature and the barrier potentials are set to $T=0.2T_c$ and 
	$z_0=0$. }
  \label{fig:ARmag_sw}
\end{figure*}
The induced magnetizations in the 2D junctions are shown in 
Fig.~\ref{fig:FSmag}, where the chemical potential in the F is set to 
(a) $\mu_F = 2\mu_S$ and 
(b) $\mu_F = -\mu_S$. When $\mu_F = 2\mu_S$, $\mathrm{sgn}(M)$ is
determined by whether the ABSs are present or not. The anomalous IPE
occurs when the ABSs are present at the interface (i.e., $p_x$- and
$d_{xy}$-wave junctions). 
When $M_0<\mu_F$, all of the channels are regarded as FM/SC junctions which
results in the parallel (antiparallel) magnetization in the SC without
(with) ABSs as discussed in the 1D limit [see Fig.~\ref{fig:magFMM}]. The
direction of the induced magnetization does not change even in the
$M_0 > \mu_S$ region. 

When $\mu_F = -\mu_S$, $\mathrm{sgn}(M)$ for each
superconducting junction is inverted compared with Fig.~\ref{fig:FSmag}(a). 
In the $M<\mu_F$ region, all of the channels are regarded as an FI/SC junction
where the induced magnetization in the absence (presence) of ABSs is
negative as discussed in the 1D limit. When $M_0 > \mu_F$, even though 
the channels around $k_y = 0$ change to HM/SC junctions, the sign of
$\thsm$ remains unchanged. Therefore, the total amplitude of the
magnetization also remains unchanged. 

When $M_0 \gg \mu_F$ and $\mu_F = \mu_S$, 
the sign change of $\thsm$ can not be ignored. 
The $M_0$ dependence of $M|_{x=0^+}$ for $\mu_F = \mu_S$ are shown in
Fig.~\ref{fig:FSmag2}. In the $p_y$- and
$d_{xy}$-wave case (i.e., SCs with  gap nodes on the Fermi
surface at $k_y=0$), the direction of the induced magnetization changes
around $M_0 =
\mu_F$ as shown in Fig.~\ref{fig:FSmag2}. On the other hand, the signs
of $M|_{x=0^+}$ for 
the $s$-, $p_x$-, and $d_{x^2-y^2}$-wave superconducting junction (i.e., SCs without 
gap does at $k_y=0$) are unchanged.

To understand the sign change of the induced magnetization, we
evaluated the angle-resolved magnetization $M(\vphi)$ with $k_y = \sin
\vphi$. The results for $s$- and $p_y$-wave junctions with $\mu_F = 2\mu_S$ 
are shown in Figs.~\ref{fig:ARmag_sw}(a) and \ref{fig:ARmag_sw}(b). In
this case, $M(\vphi)$ are positive for both of the junctions because
the sign of $\thsm$ is always positive as shown in
Fig.~\ref{fig:ARmag_sw}(c). Note that $M(\vphi)|_{\vphi=0}=0$ in the
$p_y$-wave junction because of the nodes on the superconducting gap
shown by the broken red line in Fig.~\ref{fig:ARmag_sw}(b). 

The results with $\mu_F = \mu_S$ are shown in
Figs.~\ref{fig:ARmag_sw}(d), \ref{fig:ARmag_sw}(e), and
\ref{fig:ARmag_sw}(f). The spin-mixing angle can be negative when $M_0
\neq 0$ as shown in Fig.~\ref{fig:ARmag_sw}(f) \cite{Eschrig_Royal_18}. In the $s$-wave
junction with $M_0 \neq 0$, the positive contribution (shown in green) around $\vphi =
0$ is larger than the negative ones (shown in purple) around $|\vphi| = \pi/2$. The
total magnetization, therefore, is always positive even for $\mu_F =
\mu_S$ as shown in Fig.~\ref{fig:FSmag2}. In the $p_y$-wave junction,
on the other hand, the positive contribution is smaller than in the
$s$-wave case because of the nodes. As a result, the direction of the total magnetization can change around $M_0 = \mu_F$ (Fig.~\ref{fig:FSmag2}).

\section{Discussions}

The anomalous IPE presented in this paper can be observed by 
ferromagnetic resonance (FMR) measurements \cite{FMR_1998,FMR_2002}, by 
nuclear magnetic resonance (NMR)  measurements
\cite{Salikhov_PRL_2009}, and by  polar Kerr effect measurements
\cite{Xia_PRL_2009}. In these experiments the conventional IPE has
been observed. Replacing the conventional SC by an USC such as a
high-$T_c$ cuprate, it would possible to confirm experimentally the
anomalous IPE. 

In this paper, we assume that the pair potential depends only on the temperature but not on the coordinate. 
This assumption, however, changes the results only
quantitatively but not qualitatively. Our main conclusion about 
the direction of the induced magnetization would remain unchanged 
even if we employ the spatial-dependent self-consistent pair potential.

\section{Conclusion}

We have theoretically studied the IPE in F/USC junctions utilizing the
quasiclassical Green function theory. We have shown
that the direction of the induced magnetization is determined by 
two factors: by whether the ABS exists and by the sign of the spin-mixing angle $\thsm$. 
Namely, in the 1D limit, the induced magnetizations for the 
$p_x$-wave SC is always opposite to that for the $s$-wave SC. 

In the 2D model, the spin-mixing angle $\thsm$ depends on the momentum
parallel to the interface $k_y$. The results for 2D F/SC junctions
are qualitatively the same as those in the 1D limit when the magnetization in the F ($M_0$) is smaller than the chemical potential of the F ($\mu_F$). When $M_0 \gg \mu_F$, the sign of the
induced magnetization is not simply determined by the ABSs because 
the sign changes of $\thsm$ around $k_y \sim \pm k_F$ are 
not negligible in this parameter range.

In addition, analyzing the pair amplitudes in 1D models, 
we have pointed out a
correspondence at $T \sim T_c$ between 
the spin-singlet pairs in an $s$-wave junction and the spin-triplet pairs 
in a $p$-wave junction. The odd-frequency spin-triplet
$s$-wave pairs induced in the spin-triplet $p$-wave junction have
qualitatively the same $M_0$ dependence as that for the spin-singlet
$s$-wave pairs induced in the $s$-wave junction. 
Reflecting this correspondence, the amplitudes of the induced magnetizations in the $s$- and $p$-wave junctions are qualitatively the same. Their directions,
however, are opposite to each other, where the direction of the
magnetization is determined by the relative phase between the spin-singlet and spin-singlet pair functions.

\begin{acknowledgments}
This work was supported by the JSPS Core-to-Core program ``Oxide Superspin'' international network. 
This work was supported by Grants-in-Aid from JSPS for Scientific
Research on Innovative Areas ``Topological Materials Science''
(KAKENHI Grant Numbers JP15H05851, JP15H05852, JP15H05853, 
and JP15K21717), Scientific
Research (A) (KAKENHI Grant No. JP20H00131), 
Scientific Research (B) (KAKENHI Grant Numbers JP18H01176 and JP20H01857), 
Japan-RFBR Bilateral Joint Research Projects/Seminars number 19-52-50026. S-I. S. acknowledges hospitality during his stay at University of Greifswald, Germany.
\end{acknowledgments}

\appendix


\section{Symmetry of Cooper pairs and induced magnetization}
\label{sec:appen1}

When there is a spin-dependent potential, subdominant pairing
component must be induced because of the symmetry breaking. 
Near the interface of an F/SC junction, the anomalous Green
functions are expressed as a superposition of the spin-triplet and
singlet pairs: 
\begin{align}
       \hat{g}= 
	\mathrm{diag}[ g_{\ua}, g_{\da}], \hspace{6mm}
  &     \hat{f}	= f_{0} i\hat{\sigma}_2 + f_{3} \hat{\sigma}_1, \\[0mm]
  & \ut{\hat{f}}= \ut{f}_{0} (-i\hat{\sigma}_2) + \ut{f}_{3} \hat{\sigma}_1, 
\end{align}
where we consider a spin-dependent potential parallel to the spin
quantization axis.  From the normalization condition (i.e., 
$\hat{g}^2 - \hat{f} \ut{\hat{f}}= \hat{\sigma}_0$), we have the explicit forms for
$g_{\ua}$ and $g_{\da}$:
$
  g_{\ua(\da)}^2 = 
	     [ 1 + f_0 \ut{f}_0 + f_3 \ut{f}_3 ]
	+(-) [     f_0 \ut{f}_3 + f_3 \ut{f}_0 ]
$. 
 When $T \sim T_c$, 
the pair amplitude is sufficiently small. Accordingly, the approximated
Green's function and the local magnetization are given by the following expressions:
\begin{align}
  &g_{\ua(\da)} = 
	     1 + \frac{1}{2} \left\{ [ f_0 \ut{f}_0 + f_3 \ut{f}_3 ]
	+(-) [     f_0 \ut{f}_3 + f_3 \ut{f}_0 ] \right\}
	\notag \\[3mm]
  & M(x) 
  \approx 2 \pi \mu_B N_0T \sum_{\omega_n>0} 
	\int_{-\pi}^{\pi} \mathrm{Im}\big[ f_0 \ut{f}_3 + f_3 \ut{f}_0 \big]
	\, \frac{d \vphi}{2\pi}. 
\end{align}

In a 2D system, the anomalous Green's function can be
expanded in a Fourier series: 
\begin{align}
  f_\nu = \frac{C_{\nu,0}}{\sqrt{2\pi}} + \frac{1}{\sqrt{\pi}} \sum_{l>0} 
	\left[ 
	  C_{\nu,l} \cos (l\vphi) 
	+ S_{\nu,l} \sin (l\vphi) 
	\right], 
  \label{}
\end{align}
where $C_{\nu,n}$ are $S_{\nu,n}$ are coefficients that represent
each pairing amplitude (e.g., $C_{\nu=0,0}$, $C_{3,1}$, and $S_{3,2}$ 
correspond to the $s$-wave spin-singlet, $p_x$-wave spin-triplet, 
and $d_{xy}$-wave spin-triplet pairs). 
Note that the spin-singlet odd-parity and spin-triplet even-parity
components should be odd functions with respect to the Matsubara
frequency. In other words, they represent the odd-frequency pair
amplitudes. 
Using $\ut{f}(x,\vphi,i\omega_n) =
f^*(x,\vphi+\pi,i\omega_n)$
and the orthogonality of the trigonometric functions, 
we can obtain 
\begin{widetext}
\begin{align}
  & M(x) 
  = 4 \pi \mu_B N_0 T \sum_{\omega_n>0} 
	\mathrm{Im}\big[ 
	C_{0,0}C^*_{3,0} + \sum_{l>0} 
	(-1)^l
	\left(
	  C_{0,l} C^*_{3,l} 
	+ S_{0,l} S^*_{3,l} 
	\right)
	\big]. 
\end{align}
\end{widetext}
In the 1D limit, in particular, the magnetization is given by 
\begin{align}
  & M
  = 4 \pi \mu_B N_0T \sum_{\omega_n>0} 
	\mathrm{Im}\big[ 
	  f  _{0,\mathrm{SW}}
		f^*_{3,\mathrm{SW}} 
	+ f  _{0,\mathrm{PW}} 
	  f^*_{3,\mathrm{PW}} \big], 
\end{align}
where SW and PW stand for $s$- and $p$-wave pairings (i.e.,
even-parity and odd-parity pairing in the 1D limit). The
magnetization is generated by the product of the spin-singlet and
spin-triplet pairs. 
\begin{figure}[b!]
  \vspace{0mm}
  \centering
  \includegraphics[width=0.48\textwidth]{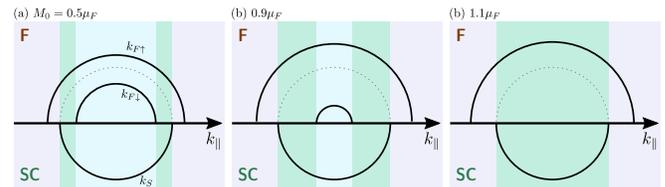}
	\caption{Evolution of the Fermi surfaces. The chemical potentials
	are set to $\mu_F=\mu_S$. The magnetization is set to 
	(a) $M_0 = 0.5\mu_F$, 
	(b) $      0.9\mu_F$, and 
	(c) $      1.1\mu_F$. 
	There are two Fermi surfaces in the F in the light-blue region,
	whereas only one Fermi surface exists in the green region. 
	The outer light-purple region is irrelevant to the IPE. 
	}
  \label{fig:FerSur}
\end{figure}

\section{Spin-mixing angle and Fermi surfaces}

The sign of the spin-mixing angle $\thsm$ is not simply determined by 
whether the F is an FM or HM \cite{Eschrig_Royal_18}. 
We show the evolution of the Fermi surfaces in
Fig.~\ref{fig:FerSur}, where the chemical potentials
are set to $\mu_F=\mu_S$. The magnetization is set to 
(a) $M_0 = 0.5\mu_F$, 
(b) $      0.9\mu_F$, and 
(c) $      1.1\mu_F$. Increasing $M_0$, the spin bands split in the F. As a
result, there is only one Fermi surface for the channels with
$k_\parallel$ as shown in the green region in Fig.~\ref{fig:FerSur}.
When $M_0>\mu_F$, the Fermi surface for the down-spin band vanishes. 
Comparing Figs.~\ref{fig:FerSur} and \ref{fig:ARmag_sw}(f), we see
that $\mathrm{sgn}[\thsm]$ is not in an obvious way related to the band structure in the F. 

In junctions of a ferromagnet and a normal metal, 
the spin-dependent potentials 
in the magnet give rise to a 
\textit{phase delay} of the wave function for 
the reflected quasiparticle \cite{Eschrig_Royal_18}. 
The quasiparticle injected from the normal metal penetrates into 
the ferromagnet even when the process is classically forbidden. The 
quasiparticle is reflected after experiencing the spin-dependent potential, 
which results in an additional phase. Therefore, the spin-mixing angle is not 
only determined by the electronic structure in the ferromagnet but by how 
the quasiparticle experiences the magnetic potentials at the interface.

\bibliography{tsc05}

\end{document}